\documentclass[fleqn,10pt]{wlscirep}
\usepackage[utf8]{inputenc}
\usepackage[T1]{fontenc}

\DeclareMathOperator{\sinc}{sinc}
\usepackage{color,amsmath,scalerel,soul}
\usepackage{moresize} 
\usepackage{bm}
\usepackage{blindtext}

\newcommand{\ket}[1]{\lvert #1 \rangle}

\newcommand{\si}{\scaleto{i}{3.8pt}}

\title{Frequency and polarization emission properties of a photon-pair source based on a photonic crystal fiber}

\author[1]{Daniel De la Torre-Robles}
\author[2]{Francisco Dominguez-Serna}
\author[3]{Gisell Lorena Osorio}
\author[4]{Alfred  B. U'Ren}
\author[1]{David Bermudez}
\author[3,*]{Karina Garay-Palmett}
\affil[1]{Departamento de F\'isica. Cinvestav, A.P. 14-740, 07000 Ciudad de M\'exico, M\'exico}
\affil[2]{C\'atedras Conacyt - Centro de Investigaci\'on Cient\'ifica y de Educaci\'on Superior de Ensenada, B.C., 22860, M\'exico}
\affil[3]{Departamento de \'Optica, Centro de Investigaci\'on Cient\'ifica y de Educaci\'on Superior de Ensenada, B.C., 22860, M\'exico}
\affil[4]{Instituto de Ciencias Nucleares, UNAM. Circuito Exterior s/n, C.U., Coyoac\'an,
C.P. 04510, Ciudad de M\'exico, M\'exico}

\affil[*]{kgaray@cicese.edu.mx}

\begin{abstract}
In this work we experimentally demonstrate a photon-pair source with correlations in the frequency and polarization degrees of freedom. We base our source on the spontaneous four-wave mixing (SFWM) process in a photonic crystal fiber.  We show theoretically that the two-photon state is the coherent superposition of up to six distinct SFWM processes, each corresponding to a distinct combination of polarizations for the four waves involved and giving rise to an energy-conserving pair of peaks.  Our experimental measurements, both in terms of single and coincidence counts,  confirm the presence of these pairs of peaks, while we also present related numerical simulations with excellent experiment-theory agreement. We explicitly show how the pump frequency and polarization may be used to effectively control the signal-idler photon-pair properties, defining which of the six processes can participate in the overall two-photon state and at which optical frequencies.  We analyze the signal-idler correlations in frequency and polarization, and in terms of fiber characterization, we input the SFWM-peak experimental data into a genetic algorithm which successfully predicts the values of the parameters that characterize the fiber cross section, as well as predict the particular SFWM process associated with a given pair of peaks.  We believe our work will help advance the exploitation of photon-pair correlations in the frequency and polarization degrees of freedom.
\end{abstract}
\begin{document}

\flushbottom
\maketitle
%
%
\thispagestyle{empty}


\section{Introduction}
Photon pair generation based on the $\chi^{(3)}$ nonlinearity holds much promise for the advancement of quantum processing technologies \cite{Wang2020}. These sources rely on the spontaneous four-wave mixing (SFWM) process \cite{fiorentino2002all}, in which two pump photons are annihilated, leading to the emission of signal and idler photon pairs. SFWM sources lead to several advantages, compared to spontaneous parametric down-conversion (SPDC) sources based on $\chi^{(2)}$ materials \cite{burnham1970observation}, such as higher conversion efficiency \cite{Fulconis:05,GarayPalmett2010} and greater flexibility in controlling the quantum correlation properties in different degrees of freedom \cite{Takesue2004,GarayPalmett2007,CruzDelgado2016}. 

SFWM photon-pair sources have been demonstrated in different types of fibers, e.g. single-mode  \cite{PARK2021126692}, weakly-guiding  birefringent \cite{GarayPalmett2016}, dispersion-shifted  \cite{Takesue:05}, few-mode \cite{Guo:19}, and photonic crystal fibers, (PCF),  \cite{Rarity:05}, as well as in integrated waveguides \cite{fiorentino2002all,Azzini:12}.  In part due to this breadth of possible source types, the SFWM process is remarkably versatile, permitting a number of source characteristics, including: i) single- or multi-mode operation (the latter permitting intra-mode phasematching) \cite{fiorentino2002all,CruzDelgado2016}, ii)  counter- or co-propagating\cite{Fulconis:05,MonroyRuz2017}, as well as cross- or co-polarized \cite{GarayPalmett2007,Fan:05} geometries, iii) the use of degenerate or non-degenerate pumps \cite{fiorentino2002all,Fan:05,Fang:13}, and iv) the presence of quantum correlations or entanglement in polarization\cite{Takesue2004_3,Zhou:09}, frequency, transverse mode \cite{CruzDelgado2016}, \cite{Li2009,Lee:06}, or time-bin\cite{Harada08} degrees of freedom.   Among the above optical fiber alternatives, it is well known that PCFs exhibit unique characteristics that make them an attractive medium for photon-pair source implementations  \cite{Rarity:05,Fan:05b,Fulconis2007}, exhibiting engineered dispersion properties leading to spectrally or temporally tailored two-photon state generation, and high effective nonlinear coefficients permitting large emission rates.

SFWM photon pair sources can be geared towards a variety of quantum information processing tasks and applications \cite{Flamini_2018}.  Among all possible quantum states, those exhibiting hyperentanglement, simultaneous entanglement in multiple degrees of freedom of a quantum system \cite{Li2018,Chen2008,ramelow2009discrete,Yu-Bo2010,Xiao-Min2021}, and hybrid entanglement, i.e. entanglement between different degrees of freedom, \cite{Neves2009,Kreis2012} have attracted recent interest. In particular, hybrid entanglement finds applications in asymmetric optical quantum networks \cite{Cavalcanti2015}, quantum key distribution protocols \cite{Nunn:13}, qubit-qudit entangled state generation \cite{Neves2009}, among others.  Photon pairs with hybrid entanglement have been proposed or demonstrated in the following degrees of freedom: path-polarization \cite{Ma2009}, frequency-spatial fiber mode \cite{Cruz-Delgado2016}, polarization-transverse spatial \cite{Nogueira2010,Bharadwaj:18}, and polarization-orbital angular momentum \cite{Nagali:10,Jabir:17,Karimi2010}. 

Here we propose and experimentally demonstrate a SFWM photon pair source, with correlations in the spectral and polarization degrees of freedom, in a commercial polarization-maintaining PCF. In our experiments, the generated two-photon state is a coherent superposition of different contributions, each resulting from a distinct SFWM process corresponding to a particular combination of polarizations of the four fields involved. We characterize our photon-pair source through frequency- and polarization-resolved measurements, which are then input into a genetic algorithm so as to determine the fiber cross-section geometry, from which in turn we may infer the fiber dispersion relation and the corresponding SFWM process. 
We believe our work may form the basis for future work on the use of polarization-frequency correlations in a variety of applications and contexts.

This paper is organized as follows: in Section \ref{theory} we describe the theory of SFWM in birefringent fibers; in Section \ref{experiment} we present details of the experimental implementation and in Section \ref{resultados} we present our results; in Section \ref{Discussion}, we discuss our experimental results and present a theoretical analysis of the quantum correlation properties of the emitted photon pairs.
Finally, in Section \ref{conclusions} we present our conclusions.

\section{Spontaneous four-wave mixing theory}\label{theory}

In this work we study the SFWM process in single-mode birefringent PCFs.  The SFWM process can occur in various polarization combinations amongst the four fields involved, where the polarization state is defined with respect to the principal fiber axes (we will refer to the polarization parallel to the slow axis as $x$, and to  the fast axis as $y$).
In the presence of a third-order electric susceptibility $\pmb{\chi}^{(3)}$, the two pump fields (which may be parallel along $x$, parallel along $y$, or orthogonally polarized),
can interact with the vacuum fluctuations, generating photon pairs, as constrained by energy conservation and phase-matching  \cite{Lin2007}. The third-order nonlinear polarization can be written as $\textbf{P}^{(3)}(\textbf{r},t)= \epsilon_0 \pmb{\chi}^{(3)} \scaleto{\vdots}{10.0pt}\textbf{E}(\textbf{r},t) \textbf{E}(\textbf{r},t)\textbf{E}(\textbf{r},t)$, where $\textbf{E}(\textbf{r},t)$ denotes the total electric field in the fiber, and it is assumed that the instantaneous electronic response dominates, while the Raman contribution is neglected \cite{Agrawal2013}. If the medium is isotropic, the $\pmb{\chi}^{(3)}$ tensor  has $21$ non-zero components  \cite{Boyd2008}, among which only four are independent $\chi^{(3)}_{xxyy}=\chi^{(3)}_{xyxy}=\chi^{(3)}_{xyyx}=\chi^{(3)}_{xxxx}/3$. Thus, $\textbf{P}^{(3)}(\textbf{r},t)$ can be expressed in terms of a single parameter $\chi^{(3)}\equiv\chi^{(3)}_{xxxx}$ \cite{lin2004vector}.

The electric field inside the nonlinear medium can be expressed as
\begin{align}\label{Et}
\textbf{E}(\textbf{r},t)&=\frac{1}{2}[(\bm{\mathcal{E}}_1 +\bm{\mathcal{E}}_2 ) \exp({-i \omega_p t)}+\bm{\mathcal{E}}_i  \exp({-i \omega_i t)}+\bm{\mathcal{E}}_s  \exp({-i \omega_s t)}]+\text{c.c.},
\end{align}

\noindent where $\text{c.c.}$ indicates the complex conjugate and  $\bm{\mathcal{E}}_{\mu}\equiv\bm{\mathcal{E}}_{\mu}(\textbf{r},t)$ is the slowly-varying amplitude of the field oscillating at the frequency $\omega_{\mu}$, with $\mu=1,2$ denoting the frequency-degenerate pump fields, for which $\omega_1=\omega_2=\omega_p$, and $\nu=i,s$ representing the idler ($\omega_i>\omega_p $), and signal ($\omega_s <\omega_p$) fields, respectively. Analogously to Eq.~(\ref{Et}), the slowly-varying part of the
third-order nonlinear polarization term oscillating at $\omega_{m}$ is given by \cite{lin2004vector}
 
 \begin{align}\label{pol}
\bm{\mathcal{P}}_m^{(3)} &= \frac{1}{2}\epsilon_0  \chi^{(3)} \Big[ \bm{\mathcal{E}}_m(\bm{\mathcal{E}}_{1}\cdot\bm{\mathcal{E}}^*_{1})+ \bm{\mathcal{E}}_{1}(\bm{\mathcal{E}}_m\cdot\bm{\mathcal{E}}^*_{1})+
 \bm{\mathcal{E}}^*_{1}(\bm{\mathcal{E}}_{1}\cdot\bm{\mathcal{E}}_m)+\bm{\mathcal{E}}_m(\bm{\mathcal{E}}_{2}\cdot\bm{\mathcal{E}}^*_{2})+ \bm{\mathcal{E}}_{2}(\bm{\mathcal{E}}_m\cdot\bm{\mathcal{E}}^*_{2})+
 \bm{\mathcal{E}}^*_{2}(\bm{\mathcal{E}}_{2}\cdot\bm{\mathcal{E}}_m)\\ \nonumber
 &+ \bm{\mathcal{E}}_{1}(\bm{\mathcal{E}}_2\cdot\bm{\mathcal{E}}^*_l)+\bm{\mathcal{E}}_{2}(\bm{\mathcal{E}}_1\cdot\bm{\mathcal{E}}^*_l)+\bm{\mathcal{E}}^*_{l}(\bm{\mathcal{E}}_1\cdot\bm{\mathcal{E}}_2) \Big],
 \end{align}

\noindent where $m,l= \{s,i\}$ $(m \neq l)$. The first six components correspond to cross-phase modulation (XPM) processes, and the last three to four-wave mixing. From Eq.~(\ref{pol}) it can then be shown that the SFWM interaction Hamiltonian \cite{loudon2000quantum} comprises six different processes related to different polarization combinations of the participating fields, each with a corresponding coefficient, in the explicit form 
\begin{align}\label{hamiltoniano}
\hat{H}(t)&\propto \epsilon_0 \chi^{(3)} \!\! \int\! dV \Big[ 3E_{1x}^{(+)}E_{2x}^{(+)}\hat{E}^{(-)}_{sx}\hat{E}^{(-)}_{{\si} x}+ 3E_{1y}^{(+)}E_{2y}^{(+)}\hat{E}^{(-)}_{sy}\hat{E}^{(-)}_{{\si}y}  +2E_{1x}^{(+)}E_{2y}^{(+)}\hat{E}^{(-)}_{sx}\hat{E}^{(-)}_{{\si}y} + 2E_{1x}^{(+)}E_{2y}^{(+)}\hat{E}^{(-)}_{sy}\hat{E}^{(-)}_{{\si}x}\\ \nonumber&+  E_{1x}^{(+)}E_{2x}^{(+)}\hat{E}^{(-)}_{sy}\hat{E}^{(-)}_{{\si}y}+E_{1y}^{(+)}E_{2y}^{(+)}\hat{E}^{(-)}_{sx}\hat{E}^{(-)}_{{\si}x} \Big],
\end{align}

\noindent where $E_{\mu q}^{(+)}\equiv E_{\mu q}^{(+)}(\mathbf{r},t)$   represents the positive-frequency component of the pump fields ($\mu=1,2$),  parallel to each of the two fiber principal axes ($q=x \,\,\mbox{or}\,\, y$), with a similar expression for the negative-frequency component.  Recall that in our convention the $x$ direction ($x$-polarization) is parallel to the slow axis, while the $y$ direction ($y$-polarization) is parallel to the fast axis \cite{GarayPalmett2010}.

From Eq.~(\ref{hamiltoniano}) and following a standard perturbative approach \cite{mandel1995optical}, it can be shown that the SFWM two-photon state is given by $\ket{\Psi}=\ket{\mbox{vac}}+\Gamma \ket{\Psi}_2$, where the two-photon component $\ket{\Psi}_2$ is the coherent superposition of the contributions from the different terms in the  Hamiltonian, each of which is related to one of the possible polarization combinations appearing in Eq. \ref{hamiltoniano}
\begin{align}
\label{state}
\ket{\Psi}_2 &= \sum_{j} \eta_{j} \gamma_{j}\sqrt{p_{j1}p_{j2}} \int \!\!d\omega_i \!\int \!\!d\omega_s  G_{j}(\omega_s,\omega_i)\hat{a}^\dagger(\omega_i; q_{j}) \hat{a}^\dagger(\omega_s; r_{j})\ket{\mbox{vac}},
\end{align}

\noindent and $\Gamma$ is a constant related to the source brightness. For each process $j$, labeled as in table \ref{tab:process}, $ \eta_{_j}$ takes the values $\{3,3,2,2,1,1\}$ according to Eq.~(\ref{hamiltoniano}), $\gamma_{_j}$ is the SFWM nonlinearity coefficient  \cite{GarayPalmett2010}, $p_{j\mu}$ is the average power of pump $\mu$, and $\hat{a}^\dagger(\omega_{\nu}, q_{j}) $ is the creation operator for the idler ($\nu=i$) and signal ($\nu=s$) modes,  characterized by frequency $\omega_{\nu}$ and polarization $q_{j}$($r_{j}$), giving rise to the Fock state $\ket{1_{\omega_{\nu}}^{q_j}}$. 

The function $G_{j}(\omega_s,\omega_i)$ in Eq.~(\ref{state}) can be written as $G_{j}(\omega_s,\omega_i)=\ell_j(\omega_s)\ell_j(\omega_i)F_j(\omega_s,\omega_i)$ where  $\ell_j(\omega_{\nu})=\omega_{\nu}^{1/2}/[n_j(\omega_{\nu})v_{gj}^{1/2}(\omega_{\nu})]$  ($\nu=i,s$), with $n_j(\omega_{\nu})$ and $v_{gj}(\omega_{\nu})$ the refractive index and the group velocity of the corresponding polarization mode, respectively. $F_{j}(\omega_s,\omega_i)$ is the joint spectral amplitude function (JSA) given by

\begin{align}
F_{j}(\omega_s,\omega_i)=\!\int\!\! d \omega \alpha (\omega) \alpha (\omega_i+\omega_s-\omega)\sinc \bigg( \frac{L}{2} \Delta k_{j}\bigg)\exp^{i \frac{L}{2}\Delta k_{j}},
\label{JS}
\end{align}

\noindent where $L$ is the fiber length, $\alpha(\omega)$ the pump spectral envelope, and $\Delta k_{j}$ represents the phase-mismatch written in terms of the propagation constant $k_{\mu}(\omega_{\mu})$ ($\mu=1,2,i,s$) as
\begin{align}\label{desempatamiento}
\Delta k_{j}(\omega,\omega_i,\omega_s)&= k_1(\omega)+k_2(\omega_i+\omega_s-\omega )-k_s(\omega_s)-k_i(\omega_i)-\phi_\text{NL},
\end{align}

\noindent which includes the nonlinear contribution $\phi_\text{NL}$ derived from self-phase and cross-phase modulation \cite{GarayPalmett2010}. Note from Eq.~(\ref{state}) that the photon-pair emission rate expected for a particular process $j$ is proportional to $|\Gamma|^2  \eta_j^2 \gamma_j^2 p_{j1}p_{j2}$.

In general, the two-photon state given by Eq.~(\ref{state}) constitutes an entangled state in the polarization and frequency degrees of freedom of the two emitted modes. As the entanglement resides in four variables (two for each emission mode), it is possible to make polarization-discriminated measurements on one of the generation modes (say, the signal mode) in order to project the conjugate mode (idler) to a subset of spectral components, each with its corresponding state of polarization. This post-selection process can be realized with the help of a polarizer on the path of the signal photon, oriented a an angle $\theta_{s}$ relative to the slow axis $x$. This operation can be represented by the following projection operator acting on the two-photon state in Eq.~(\ref{state})

\begin{align}\label{projector}
\hat{\pi}_s(\theta_s)&=\cos^2\theta_{\nu} \ket{1^x_{\omega_s}}  \langle 1^x_{\omega_s}| +\sin^2\theta_{\nu} \ket{1^y_{\omega_s}}  \langle 1^y_{\omega_s}| +\cos\theta_{\nu}\sin\theta_s\big(\ket{1^x_{\omega_s}}  \langle 1^y_{\omega_s}| +\ket{1^y_{\omega_s}}  \langle 1^x_{\omega_s}| \big ),
\end{align}

\noindent from which, by tracing the result over the idler-polarization and signal frequency degrees of freedom, we obtain the joint probability $P(\theta_s, \omega_i)$ of detecting an idler photon at frequency $\omega_i$, conditioned on the detection of a signal photon with a polarization determined by the polarizer orientation-angle $\theta_s$. Similarly, the joint probability $P(\theta_i, \omega_s)$ can be obtaining by interchanging the roles of the signal and idler modes. 

\section{Experimental implementation}\label{experiment}
\begin{figure}[t]
	\centering
	\includegraphics[width=0.8\textwidth]{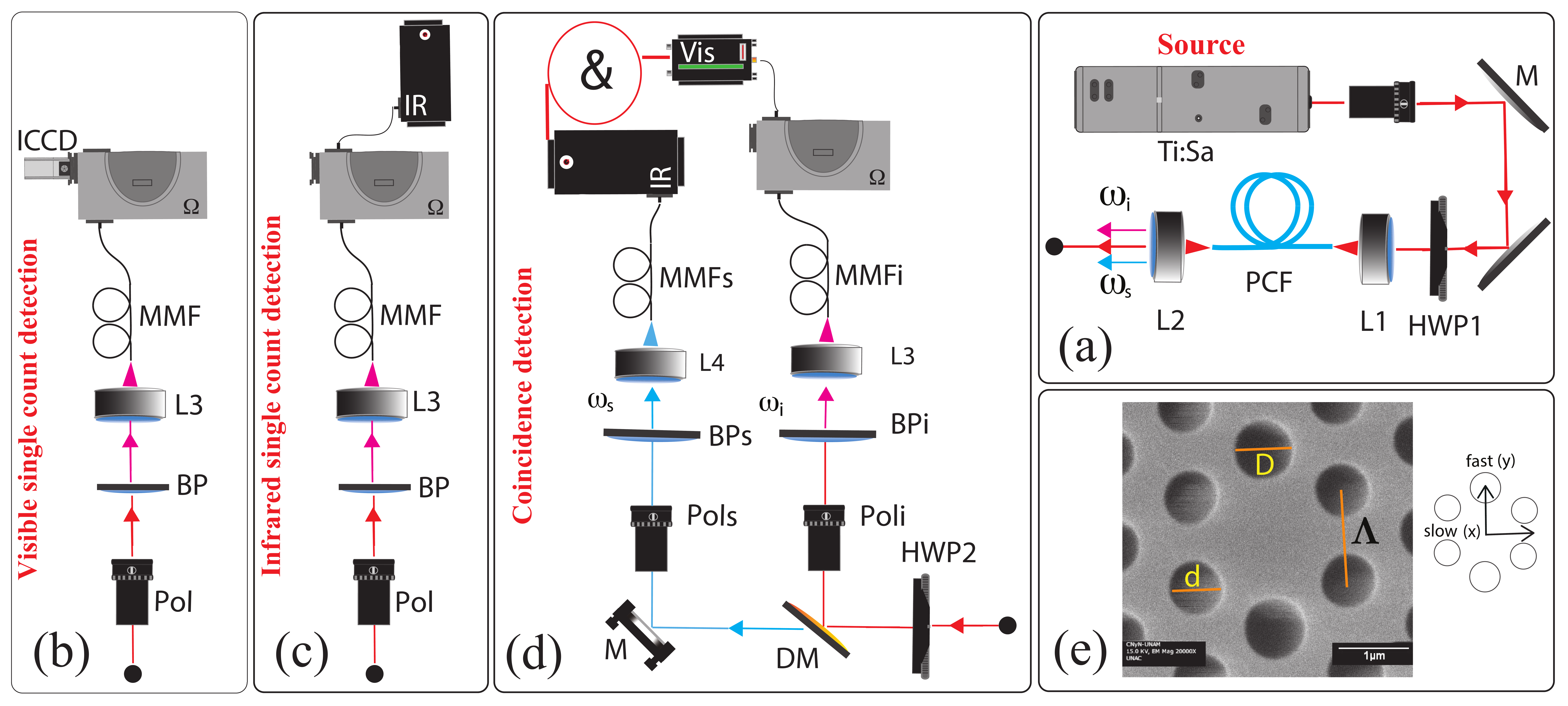}
	\caption{(Color online). Experimental setup. SFWM photon pair source implemented in a polarization maintaining photonic crystal fiber (a). Spectrally-resolved photon counting scheme with spectral resolution for idler (visible) (b) and signal (infrared) modes (c). Spectrally-resolved coincidence-photon counting scheme, showing spectral resolution for the idler mode (d). Scanning electron microscope (SEM) micrograph of the fiber profile; for the chosen reference system the slow axis coincides with the horizontal axis, while the fast axis corresponds to the vertical axis (e).}
	\label{setup1}
\end{figure}

Our photon-pair source is depicted in Figure \ref{setup1}(a). The nonlinear medium is a NL-PM-750 polarization-maintaining PCF from NKT Photonics, with a length of $0.94$m. Pump pulses are delivered from a picosecond mode-locked Ti:sapphire (Ti:sa) laser with a repetition rate of 76 MHz, tunable between 0.70 and 0.98 $\mu$m and with a full-width at half-maximum (FWHM) bandwidth of  $\sim0.4$ nm. This source configuration leads to the generation of frequency non-degenerate photon pairs. A half-wave plate (HWP1) is used to control the polarization angle of the pump light launched into the fiber. Fiber coupling is accomplished through a 42$\times$ microscope objective (L1). The idler ($\lambda_i< \lambda_p$;  in the visible) and signal $(\lambda_s >\lambda_p$; in the IR)  photons  are out-coupled from the fiber with an aspherical lens (L2).

In order to characterize the spectral and polarization properties of the signal and idler photon pairs, we first use the 
single-channel (i.e. not in coincidence) detection schemes shown in Figure \ref{setup1}(b) and (c), for the idler (visible) and for the signal (IR) emission modes, respectively. Photons emerging from L2 are transmitted through a Glan-Thompson polarizer (Pol), followed by a  bandpass filter (BP) for pump suppression, and subsequently coupled into a multimode fiber (MMF) through the lens L3, leading to a scanning grating-based monochromator $\Omega$. For  detection of the  visible idler photon, emerging from the monochromator, we use an intensified CCD camera (ICCD; iStar DH334T). For the infrared signal photon we use an InGaAs single-photon detector (IR, ID230 from ID Quantique) with an adjustable dead time in the range  2-100 $\mu$s, where the dark count rate (DKR) decreases as the DT increases. In our experiments, the ID230 DT is set to 10 $\mu$s to obtain the best trade-off between DKR and DT.  As a mechanism for characterizing the fiber-dispersion and the different SFWM interactions, spectrally-resolved single-photon counts in the idler channel are registered as a function of the pump wavelength. 

\begin{table}[t]
\centering
\caption{\bf List of the SFWM processes related to different polarization combinations of the participating electric fields.}
\begin{tabular}{ccccc}
\hline 
Process &$p_1$& $p_2$ & s \,($\lambda_s>\lambda_p$) & i \,($\lambda_p>\lambda_i$)\\
\hline
a &x&x & x& x\\
b &y&y & y&y\\
c &x&y& x& y\\
d &x&y& y& x\\
e &x&x& y& y\\
f &y&y& x& x\\
\hline
\end{tabular}
  \label{tab:process}
\end{table}

Besides the single-channel detection setup explained in the previous paragraph, we use a frequency-resolved coincidence detection system,
shown in Figure \ref{setup1}(d), to further characterize our photon pairs.
As depicted in the figure, photons emerging  from L2 are split into two paths using a dichroic mirror (DM), which reflects wavelengths $\lambda<900$nm and transmits wavelengths $\lambda>990$nm.
In each resulting arm, we place a Glan-Thompson polarizer (Pols and Poli), as well as a bandpass filter (BPs and BPi), allowing us to filter the signal and idler modes  by polarization and frequency, and finally each mode is coupled into multi-mode fibers (MMFi and MMFs for the idler and signal modes, respectively). 
In order to measure the idler spectrum in coincidence with the corresponding (spectrally unresolved) signal photons, the fiber MMFi is connected to the input port of the monochromator for frequency-resolved detection using the SPCM-AQRH (Vis, placed at the monochromator's exit), while photons coupled into fiber MMFs are delivered directly to the InGaAs detector. 
Then, coincident detection events in the two avalanche photodiodes are registered as a function of the central frequency of the monochromator's transmission window, with the temporal coincidence window set to $2.6$ ns. 
Note that we estimate the accidental coincidence count level by introducing a temporal delay, equal to the temporal separation between two subsequent pump pulses, between the signal and idler photons and monitoring the resulting count rate.  
The spectrally-resolved count rate for signal photons is measured by reversing the role of the two emission modes.
From the frequency-resolved photon counting measurements, we find that for a wide interval of pump wavelengths, accessible from our Ti:sapphire laser,  the PCF used as source fulfills phase-matching for all six SFWM processes comprising the two-photon quantum state given by Eq.~(\ref{state}).   Let us note that the polarization-maintaining property of the fiber (i.e., the ability to preserve the polarization state of light polarized along the fiber principal axes) ensures the integrity of each of the six processes along the entire length of fiber, making it possible to discriminate between them by polarization at the fiber's output. 
\begin{figure}[t]
	\centering
	\includegraphics[width=0.47 \textwidth]{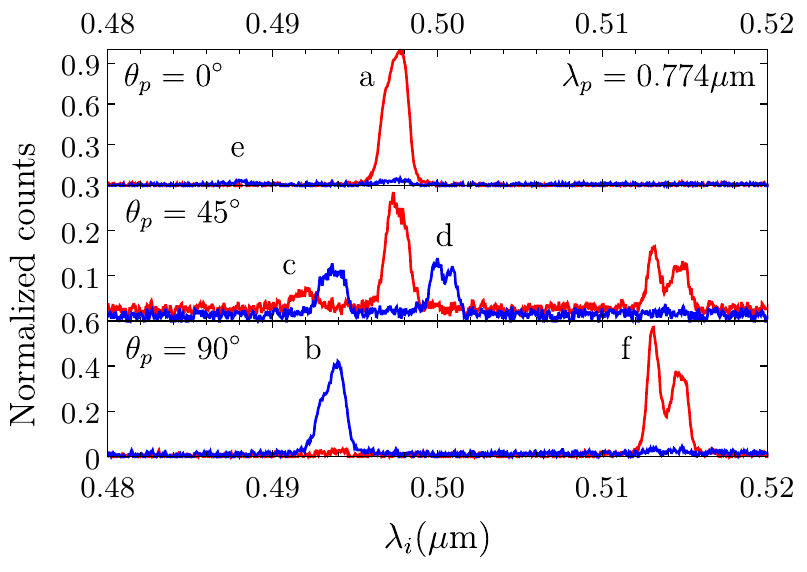}
	\caption{(Color online).  Spectrally-resolved single photon counting for the idler channel. Red and blue lines correspond to the $x$ and $y$ polarizations, respectively. From top to bottom the corresponding pump polarization angles are $\theta_p =0^o$, $\theta_p =45^o$, and $\theta_p =90^o$.}
	\label{SpecAngle}
\end{figure}

In our experiment,  the pump polarization angle can be adjusted with the help of half wave plate HWP1  so that it matches either the slow or the fast axis, or any other direction.
Thus, if the pump field is linearly polarized parallel to the slow axis, both SFWM pump photons will be $x$-polarized. 
Consequently, emitted photons must be either both $x$-polarized (this corresponds to process $a$) or both $y$ polarized  (process $e$), see Table \ref{tab:process}.
In contrast, if the pump field is linearly polarized along the fast axis, then both SFWM pump photons will be $y$-polarized. 
In this case, the generated photons must, also, either both be $y$-polarized (process $b$), or both  $x$-polarized (process $f$).   
If the pump polarization angle is set to an intermediate angle, e.g. 45$^\circ$ (angularly equidistant from the two principal axes), then it is possible to excite processes involving an $x$-polarized pump and a $y$-polarized pump (processes $c$ and $d$).
Finally, a polarization-resolved measurement of the signal and idler photons makes it then possible to discern between these pairs of processes ($a / e$, $b / f$, $c/d$).   Note that because phasematching is polarization-dependent, in general the six different processes are likely to be spectrally distinct, aiding their experimental characterization.

\section{Results}\label{resultados}

\begin{figure}[t]
	\centering
	\includegraphics[width=0.38\textwidth]{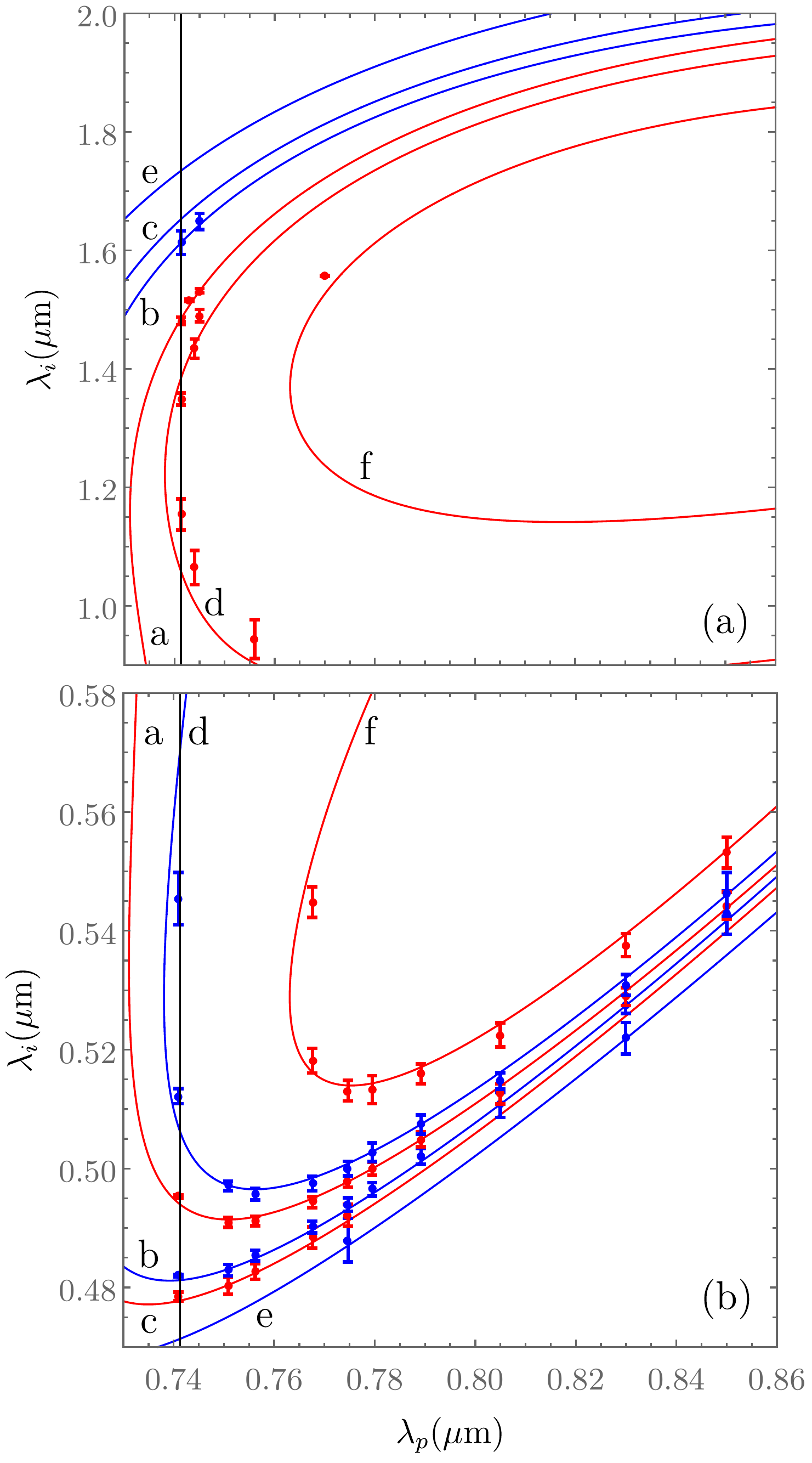}
	\caption{(Color online). Phase-matching diagrams for the six SFWM processes with signal (a) and idler (b) arms displayed separately. Red and blue colors corresponds to $x$- and $y$-polarized signals, respectively. Points indicate the central wavelength of the measured spectra, while error bars represent their emission bandwidth. Solid lines correspond to the calculated phase-matching diagrams of each process. The vertical black line represents the pump at $0.741\,\mu$m and corresponds to the spectra shown in Figure \ref{Spectrum}.}
	\label{pmcurves}
\end{figure}

From the spectrally- and polarization-resolved single-channel detection scheme, described in Section \ref{experiment}, we have obtained a set of single-photon spectra for the idler mode, as a function of the pump wavelength. In Figure \ref{SpecAngle} we have shown an example of such data, for a pump wavelength of 
 $\lambda_p=0.774\,\mu$m.   Red curves denote the $x$ polarization for the SFWM idler photon, while blue curves denote the $y$ polarization; this polarization-based discrimination is accomplished with the help of polarizer $POLi$. In conjunction with our knowledge of the pump polarization, we are then able to identify each of the spectral peaks with a particular process, $a$ through $f$; see table \ref{tab:process}. Note that some processes produce a double spectral peak, which is in fact well explained below in terms of the curvature of the underlying phasematching curves. Figure \ref{SpecAngle} has three separate panels: in the top one we show the case where we set the pump polarization angle to $\theta_p=0^\circ$ ($x$ polarization), resulting in the appearance of peaks related to processes $a$ and $e$;  the bottom one shows the case where we set the pump polarization angle to $\theta_p=90^\circ$ ($y$ polarization),  with  processes $b$ and $f$ appearing; and in the middle one we show the case where we set the pump polarization angle to $\theta_p=45^\circ$ so that processes $c$ and $d$ appear, in addition to the other four processes at a reduced emission rate since the required pump polarization state is present.

\begin{figure}[t]
	\centering
	\includegraphics[width=0.42 \textwidth]{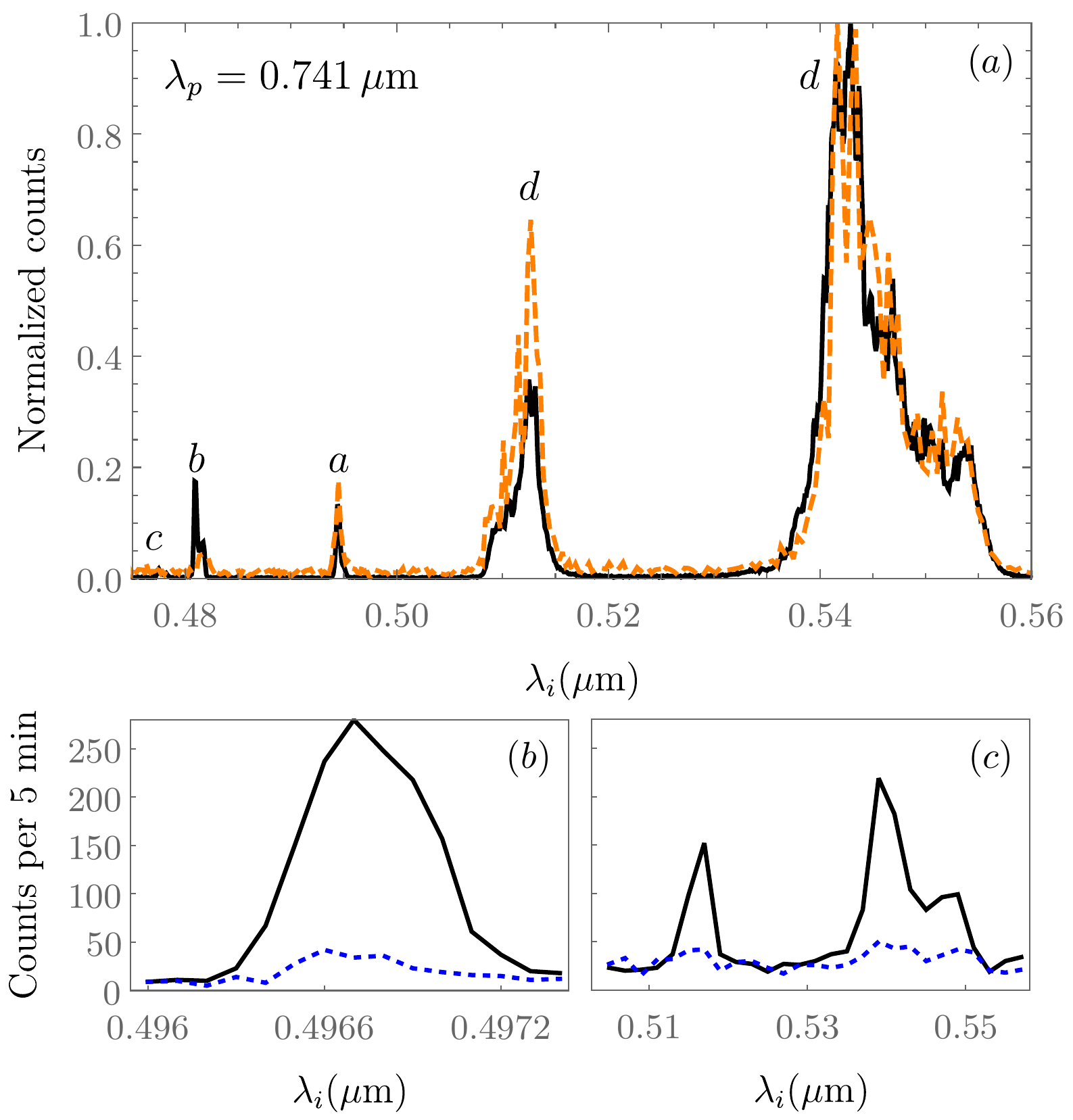}
	\caption{(Color online). (a) Single-channel spectra for $\lambda_p=0.741\,\mu$m. The solid black line is the idler mode spectrum, while the dashed orange line represents the signal field, which we have mapped to the visible range through energy conservation. (b) and (c) Idler-channel spectrally-resolved coincidence photon counting for processes $a$ and $d$, respectively. }
	\label{Spectrum}
\end{figure}

We repeated the process described above for $10$ different pump wavelengths spanning the interval from $0.741\mu$m to $0.850\mu$m. 
Our results are summarized in Figure \ref{pmcurves}, in which points indicate the central wavelength of the measured spectral peak in question, and the error bars represent the emission bandwidth. While for some of the pump wavelengths we were able to identify all six peaks, for some other pump wavelengths either not all six peaks exist within our SFWM-measurement spectral window, or the peaks could not be resolved from the noise level (this is the case in several instances for process in $e$, which leads to comparatively weaker peaks).   Also note that fewer data points were acquired for the signal (IR), as compared to the idler (visible). 

As will be discussed in the next section, the information obtained from the idler spectrum measurements is used to infer the fiber transverse geometry from which it is possible to predict the dispersion relations for each of the two polarizations $k_x(\omega)$ and $k_y(\omega)$, and in turn plot phasematching contours given by  the condition $\Delta k_j=0$ [see Eq.~(\ref{desempatamiento})] for process $j$, with $j=1$ through $6$. The resulting phase-matching diagrams are shown in Figure \ref{pmcurves} as solid lines, in which signal (panel a) and idler (panel b) arms are displayed separately. Note that in these figures, red and blue colors corresponds to $x$-polarized and $y$-polarized signals, respectively, as in Figure \ref{SpecAngle}.  Note that in some of the curves (particularly for process $d$), the phasematching contour arches in such a manner that two signal-mode peaks and two idler-mode peaks may appear for a single pump frequency within the signal/idler spectral windows used in our experiment.  Note also that the experimental data points agree remarkably well with the phasematching curves.    Because the dispersion relations for the two polarizations are different, we expect the phasematching curves for different processes (each involving a different combination of polarizations) to be spectrally distinct, as in fact is observed in our measurements. 

In Figure \ref{Spectrum}(a), we show (black curve)  the spectrally-resolved single-channel counts for the idler mode, obtained for a different choice of pump central frequency, $\lambda_p=0.741\,\mu$m, with a polarization angle $\alpha=45^o$, so that all processes can appear for a single pump polarization setting. In this case, phase-matching is fulfilled for processes $a$ through $e$, as indicated by the black vertical line in figure \ref{pmcurves}.  Note that the visible and IR peaks for the comparatively weaker process $e$, cannot be observed at the noise level in our measurement. The solid black line represents the directly-measured idler mode spectrum, and the dashed orange line represents the experimentally-measured IR signal-mode spectrum mapped to the visible under the transformation $\omega_s=2\omega_p-\omega_i$, as required by energy conservation.   The remarkable overlap between the two curves is an experimental confirmation of the expected signal-idler spectral correlations. In order to further verify the photon-pair nature of the optical flux produced by the fiber, we have detected the idler photon with spectral-resolution for processes $a$ and $d$ \emph{in coincidence} with the signal-photon left spectrally-unresolved. The results are shown in Figures \ref{Spectrum}(b) and \ref{Spectrum}(c), for processes $a$ and $d$, respectively, with black lines corresponding to the coincidence count rate, and blue lines representing the accidental coincidence count rate. Note that  we have used polarizers in the signal and idler channels of the detection system ($POLs$ and $POLi$ in Figure \ref{setup1}(b)) to avoid coincidence counts between signal and idler photons corresponding to different processes. As can be appreciated form this data, the  coincidence detection rate bridging the visible and IR channels is low, which can be understood in terms of the unmatched time response of the different detectors used  in each channel (the InGaAs APD has a deadtime three orders of magnitude higher than the silicon APD), the low transmittance of the monochromator  ($\sim2\%$), as well as the pump power kept intentionally low ($< 1$mW) to reduce the accidental coincidence count level.  

\section{Discussion}\label{Discussion}

\subsection{Fiber and SFWM process characterization}

As we have explored in the past \cite{GarayPalmett2016}, experimentally-measured SFWM spectral peaks can be used as input for a computational method based on a genetic algorithm designed to calculate the underlying fiber geometry.
In the present case, the fiber supports only the fundamental, quasi linearly-polarized modes with $x$ and $y$ polarizations, i.e.  $HE_{11x}$ and $HE_{11y}$ \cite{Yariv2007}.    In our present strategy, the parameter space,  formed by  the air-hole diameters $d$ and $D$ ($d<D$), as well as the  pitch size $\Lambda$ [see scanning electron microscope (SEM) image in  Figure \ref{setup1} (d)] is probed by the genetic algorithm.   At each particular point of this space,  a numerical finite-difference time-domain vectorial mode solver
is used to obtain the corresponding dispersion relations $k_x(\omega)$ and $k_y(\omega)$.   

We use as input for our genetic algorithm the idler-arm experimental data shown in Fig. \ref{pmcurves}(b), i.e. frequencies and polarizations of emission for each of the  six processes (see table \ref{tab:process}; each one labelled with index $j$).  Specifically, this data includes an experimentally-determined  idler-mode emission frequency $\omega_i$  for each of a collection of pump frequencies $\omega_p$, labelled with index $i$.  Note that the corresponding signal frequency is inferred through energy conservation as $\omega_s=2 \omega_p-\omega_i$; the signal-photon spectra contain duplicated information, see Fig. \ref{pmcurves}, and can therefore be excluded from the genetic algorithm.  We can then evaluate the phasemismatch $\Delta k_{ij}$ at the the frequencies $\omega_p$, $\omega_i$, and $\omega_s$ for each process $j$ and for each pump frequency $i$ through Eq.~(\ref{desempatamiento}). We use as fitness function (FF) in our genetic algorithm $\Delta k_T(d,D,\Lambda)=\sum_i\sum_j\Delta k_{ij}^2$, which takes into account the collective behavior across the 6 processes and all pump frequencies considered.




The genetic algorithm not only determines the best numerical values for the parameters $d$, $D$, and $\Lambda$, but also the polarizations of the two pump waves and the signal, and consequently  determines the particular SFWM process (from those listed in Table \ref{tab:process}) responsible for each idler peak in question.  From our genetic algorithm-powered fiber characterization method, we have obtained the following parameters that minimize the FF:  $d=0.702\,\mu m$, $D=0.820\,\mu m$ and $\Lambda=1.088\,\mu m$. Note that these agree well with the values obtained directly from the SEM image shown in Figure \ref{setup1}(e): $d=0.719 \pm 0.041 \mu$m, $D=0.824 \pm 0.045 \mu$m, and $\Lambda=1.147 \pm 0.065 \mu$m.
Based on the values from the genetic algorithm, we have plotted in Figure \ref{pmcurves} the phasematching contours (composed of pump and SFWM
frequencies fulfilling perfect phasematching) for all six processes detailed in Table \ref{tab:process}.  
We employ the same color code as for the experimental data, red and blue colors represent the $x$ and  $y$ polarizations, respectively. It can be appreciated in the figure that there is excellent general agreement between the experimental data points and the calculated phasematching contours. Hence, our approach which involves inputting idler spectra originating from different central pump wavelengths into the genetic algorithm permits us to characterize the fiber dispersion for the two orthogonally-polarized $HE_{11x}$ and $HE_{11y}$ fundamental modes, within a considerably wide spectral window.

Note that the PCF on which we have based these experiments is characterized by a large dielectric contrast and a small core radius, and exhibits some unique dispersion properties: a reduced anomalous dispersion spectral window, blue-shifted to the visible. 
This leads to closed-loop phase-matching contours for co-polarized processes (such as processes $a$ and $b$), allowing us to bridge the visible and infrared regions in our photon pair emission \cite{Soeller2010}. 
Pumping in the normal dispersion region allows the generation of photon pairs, lying on the external phase-matching contours (see Ref. \cite{Fulconis:05}),  well separated from the pump and free from Raman contamination, as shown in Figure \ref{pmcurves} \cite{GarayPalmett2011b}. 

\begin{figure}[t]
	\centering
	\includegraphics[width=0.49\textwidth]{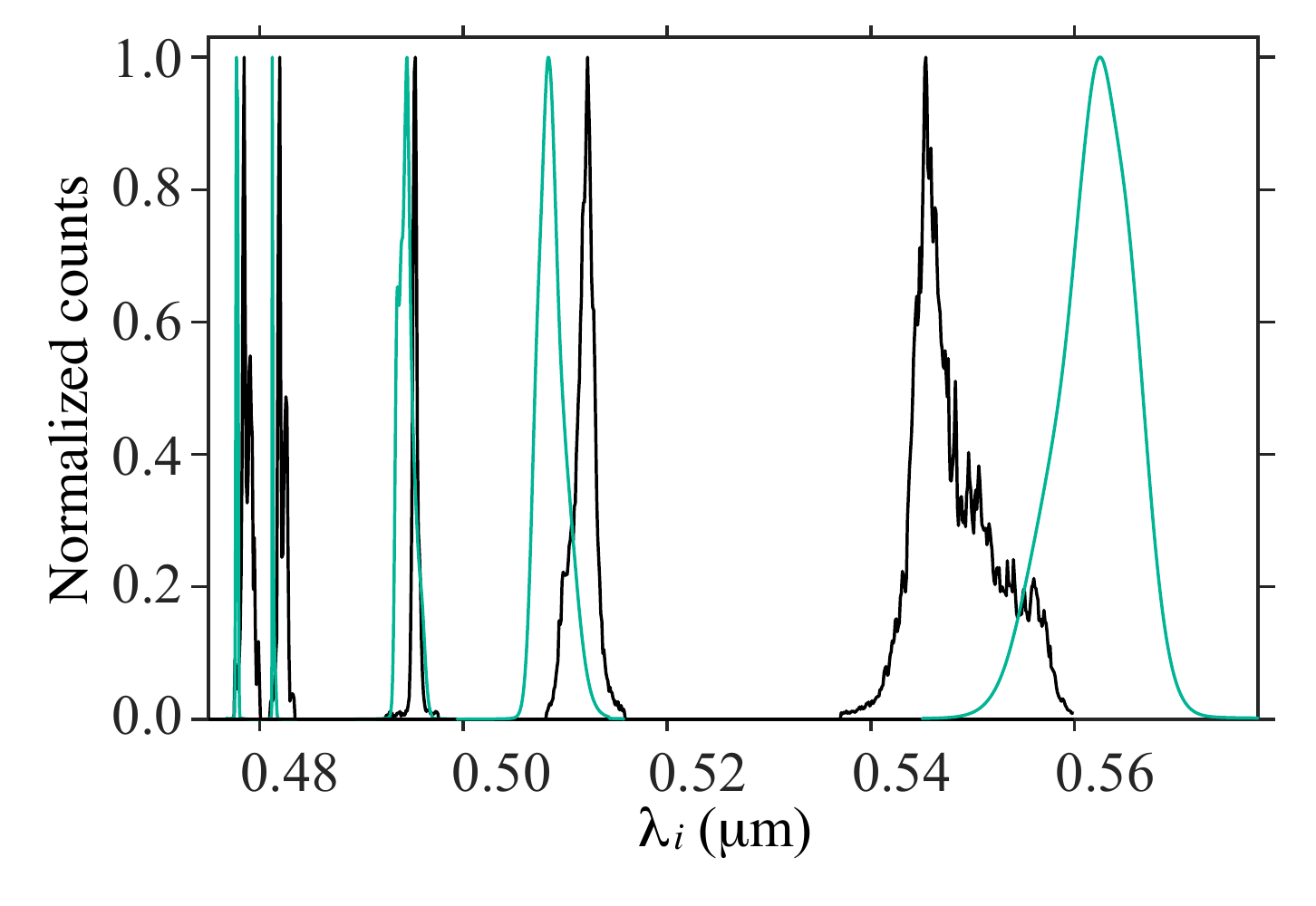}
	\caption{(Color online). Comparison between normalized experimental (black) and numerically-simulated spectra (green) for $\lambda_p=0.741\,\mu$m.}
	\label{exptheo}
\end{figure}

As a further test of the validity of our approach
we have simulated the single-channel idler spectra emitted by each of the six SFWM processes (see Table \ref{tab:process}), based on the dispersion relation derived from the fiber geometry identified by the genetic algorithm. 
We calculate these spectra from Eq.~(\ref{JS}) and the expression $I_j(\omega_i)=\int d\omega_s |F_j(\omega_s,\omega_i)|^2$ \cite{GarayPalmett2010}, for the marginal idler-mode spectrum, using  our experimental parameters: fiber length $L=0.94$ cm, pump centered at $\lambda_p=0.741\,\mu$m with bandwidth $\Delta\lambda_p=0.4$ nm, coupled average power $p_{avg}=20$mW, and repetition rate $f_r=76$ MHz. Our results are shown in Figure \ref{exptheo} (gray lines), normalized so that all peaks have the same height.   For comparison, black curves show the similarly normalized experimental results; it is clear that there is a broad agreement between the simulations and measurements.
Note that in our simulation of the SFWM spectra, we have taken into consideration the propagation effects undergone by the pump pulse, of which self- and cross-phase modulation processes are of significant importance because of the large nonlinear coefficient that characterizes our particular fiber \cite{Agrawal2008}. 

\subsection{State preparation, as controlled by the pump}

The SFWM pump in our experiments is defined by two main characteristics: its optical frequency and its polarization. On the one hand, it is clear from Figure \ref{SpecAngle}, and from Table \ref{tab:process}, that the pump polarization controls which of the six processes may occur.  On the other hand, it is clear from Figure \ref{pmcurves}
that by spectrally tuning the pump frequency, one can influence the resulting signal and idler frequencies according to the phasemartching contours.   Note that certain pump frequencies lead to a dual peak for each of the signal and idler (as is indeed shown for the specific case $\lambda_p=0.741\mu$m).

This dependence of the emission characteristics on the pump polarization and frequency may be used as a means to control the two photon state.

\subsection{Quantum correlations in the two-photon state}

In this section, we illustrate through simulations the spectral-polarization quantum correlation properties of the two-photon state emitted from our SFWM source. 

\begin{figure}[t]
	\centering
	\includegraphics[width=0.44\textwidth]{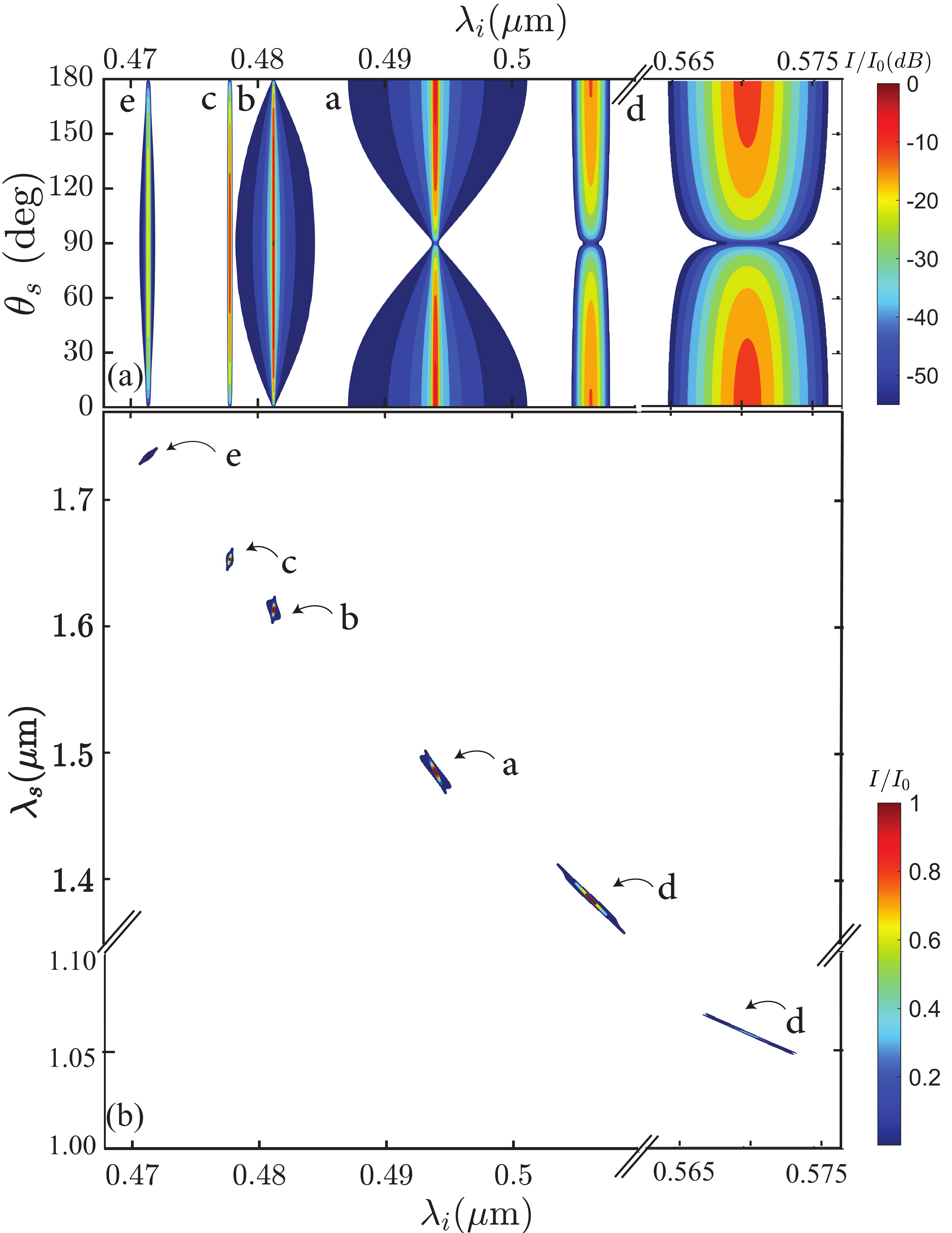}
	\caption{(Color online). For a pump field centered at $0.741\,\mu$m: (a) Joint probability density $P(\theta_s, \omega_i)$ expressed in decibels and (b) normalized joint spectral intensity of the two-photon state, plotted as filled contours.}
	\label{corr}
\end{figure}

\begin{figure*}[t]
	\centering
	\includegraphics[width=0.78\textwidth]{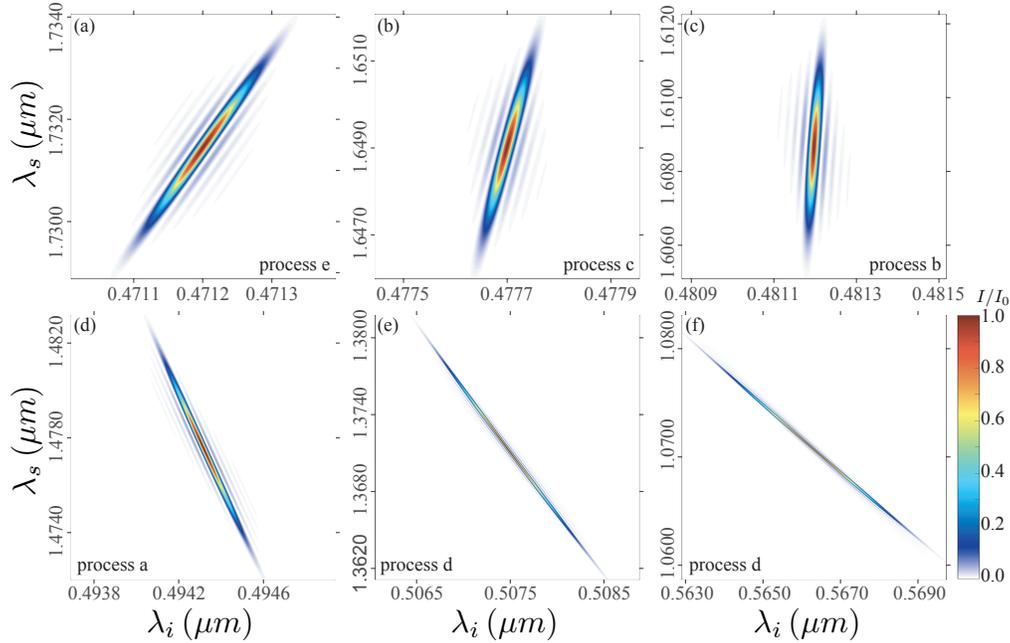}
	\caption{(Color online). Normalized joint spectral intensity of the individual processes shown in figure \ref{corr}.}
	\label{JSIs}
\end{figure*}

In Figure \ref{corr}(b) we plot the joint spectral intensity (JSI) for each of the six processes ($a$ through $f$), for a particular case involving a $0.94\,\mu$m fiber length and a pulsed pump centered at $\lambda_p=0.741\,\mu$m with $0.4$ nm bandwidth (matching our experimental settings). While the JSI is computed for each process separately, all six resulting sub-states are plotted together. Note that in this example phasematching in not attained for process $f$ within the spectral window of interest.  These plots were obtained by tracing out the polarization for the signal and idler modes.  Because the six processes correspond to different combinations of polarizations for the three waves involved, and because the dispersion relations are polarization-dependent, the frequencies at which phasematching is attained for these six processes tend to be distinct, i.e. non-overlapping.  The degree of spectral correlation exhibited by each individual  sub-state varies considerably from anticorrelated (processes $a$ and $d$), to positively correlated (processes $b$, $c$ and $e$). In the case of process $d$, the two-photon state is composed of two distinct spectral zones, resulting from two separate solutions to $\Delta k=0$ in Figure \ref{pmcurves} that exist for certain pump wavelengths. For greater clarity, we have plotted each of the sub-sates separately in Figure \ref{JSIs} in axes involving the same spectral range for the signal and idler modes.  As explored in reference \cite{GarayPalmett2007}, the angular orientation of the spectral correlations $\theta_{si}$ (in $\{\omega_s,\omega_i\}$ space) are determined by the angular orientation of the phasematching contours $\theta_{pm}$ (in $\omega_p,\omega_s-\omega_p$  space) according to the relationship $\theta_{pm}=45^{\circ}-\theta_{si}$ \cite{GarayPalmett2007}.

We now turn our attention to the frequency-polarization correlation exhibited by our two-photon source. For this purpose, let us consider a measurement in which we are able to resolve only the signal-mode polarization and the idler-mode frequency.      We proceed to calculate the joint probability density $P(\theta_s, \omega_i)$ by acting the projective operator given by Eq.~(\ref{projector}) on the two-photon state in Eq.~(\ref{state}), while tracing out the remaining two variables $\theta_i$ and $\omega_s$.  A plot of $P(\theta_s, \omega_i)$ is shown in Figure \ref{corr}(a) assuming that the pump is centered at $0.741\,\mu$m with a polarization angle of $45^{\circ}$ (so that equal-magnitude pump components are present parallel to both the fast and slow axes, thus enabling all six processes).   Note that each individual process does not exhibit polarization-frequency correlation, i.e given the detection of a photon at frequency $\omega_i$, it is unfeasible to determine the corresponding $\theta_s$ value.  However the overall two-photon state presents correlation between $\omega_i$ and $\theta_s$, in the sense that post-selecting a particular value for $\theta_s$ leads to certain specific idler spectral zones of emission.  For example, detection of a signal photon with $\theta_s = 0^{\circ}$ ($\theta_s = 90^{\circ}$), corresponds to the idler-photon emission bands for processes $a$ and $d$ ($b$, $c$, and $e$), while detection of a signal photon with $\theta_s = 45^{\circ}$  corresponds to idler-mode emission bands corresponding to all six processes.

Finally, let us point out that the correlations in frequency and polarization discussed in this paper open up the possibility for hybrid entanglement being present in these degrees of freedom.  Particularly, the two-photon state in Eq. (\ref{state}) can indeed exhibit hybrid entanglement, for example in the signal polarization and idler frequency,  as revealed by tracing over the two remaining variables.  Such entanglement depends on the  overlap between  the joint spectra corresponding to different processes (out of those listed in table \ref{tab:process}). We hope to further explore this polarization-frequency hybrid entanglement in a future work.

\section{Conclusions}\label{conclusions}

We have demonstrated a photon-pair source, exhibiting correlations in the polarization and spectral degrees of freedom, based on the spontaneous four-wave mixing (SFWM) process in a polarization-maintaining photonic crystal fiber. As we have described theoretically, the two-photon state produced by our source is in the form of a coherent superposition of up to six distinct SFWM processes, each defined by a particular combination of polarizations for the four participating waves.  We have experimentally characterized how the pump polarization and frequency control which of these six possible SFWM processes appear in the overall two-photon state. We have presented frequency-resolved measurements of the signal (in the IR) and idler (in the visible) emission modes,  which are in the form of energy-conserving peaks, one pair of peaks per participating process. We present such measurements both in terms of single and coincidence events. Based on a theory that includes the effects of self- and cross-phase modulation of the pump waves, we present simulations of the idler-mode emission peaks, showing good agreement with the experimental data. We also show an analysis of the correlations in the frequency and polarization degrees of freedom. From our photon-pair spectrally-resolved measurements and relying on a genetic algorithm, we are able on the one hand to characterize the PCF in terms of the parameters that define its cross-section, and on the other hand to predict to which of the six processes each pair of peaks is associated.
We believe our experimental and theoretical results will contribute to future demonstrations of fiber-based or waveguide-based photon-pair sources exhibiting frequency-polarization correlations.


\bibliography{Refs}

\section*{Funding Information}
Consejo Nacional de Ciencia y Tecnología (CONACYT) (grants Laboratorios Nacionales 314860/2020, Ciencias de Frontera 51458/2019, and Cátedras-Conacyt 709/2018).

\section*{Acknowledgements}

The authors thank Israel Gradilla from CNyN (UNAM) for technical support in acquiring the SEM images of the fiber cross-section.  

\section*{Author contributions statement}

D.D.L.T-R. and F.D-S. conducted the experiments, K. G-P., D.B., F.D-S. and G.O. performed the simulations, D.B., A.U. and K.G-P. analysed the results. All authors reviewed the manuscript.

\end{document}